\documentclass[prb,preprint,]{revtex4-1}

\usepackage{graphicx}
\usepackage{color}

\begin{document}

\begin{center}
\textbf{\large Direct characterization of photo-induced lattice dynamics \\in BaFe$_2$As$_2$}\\
\vspace{10mm}

S.~Gerber$^1$, K.~W.~Kim$^{2,*}$, Y.~Zhang$^{1,3}$, D.~Zhu$^4$, N.~Plonka$^{1,5}$, M.~Yi$^{1,5}$, G.~L.~Dakovski$^4$, D.~Leuenberger$^1$, P.~S.~Kirchmann$^1$, R.~G.~Moore$^1$, M.~Chollet$^4$, J.~M.~Glownia$^4$, Y.~Feng$^4$, J.-S.~Lee$^6$, A.~Mehta$^6$, A.~F.~Kemper$^7$, T.~Wolf$^8$, Y.-D.~Chuang$^3$, Z.~Hussain$^3$, C.-C.~Kao$^9$, B.~Moritz$^1$, Z.-X.~Shen$^{1,5,*}$, T.~P.~Devereaux$^{1,*}$ \& W.-S.~Lee$^{1,*}$\\

\vspace{10mm}\textit{\small
$^{1}$Stanford Institute for Materials and Energy Sciences, SLAC National Accelerator Laboratory, Menlo Park, California 94025, USA\\
$^{2}$Department of Physics, Chungbuk National University, Cheongju 361-763, Korea\\
$^{3}$Advanced Light Source, Lawrence Berkeley National Laboratory, Berkeley, \\California 94720, USA\\
$^{4}$Linac Coherent Light Source, SLAC National Accelerator Laboratory, Menlo Park, \\California 94025, USA\\
$^{5}$Departments of Physics and Applied Physics, Stanford University, Stanford, 
\\California 94305, USA\\
$^{6}$Stanford Synchrotron Radiation Lightsource, SLAC National Accelerator Laboratory, \\Menlo Park, California 94025, USA\\
$^{7}$Computational Research Division, Lawrence Berkeley National Laboratory, Berkeley, \\California 94720, USA\\
$^{8}$Institute for Solid State Physics, Karlsruhe Institute of Technology, 76021 Karlsruhe, Germany\\
$^{9}$SLAC National Accelerator Laboratory, Menlo Park, California 94025, USA\\}

\vspace{10mm}
{\small$^*$To whom correspondence should be addressed: \textit{leews@stanford.edu, kyungwan.kim@gmail.com, tpd@stanford.edu} and \textit{zxshen@stanford.edu}.}
\end{center}

\clearpage

\noindent\textbf{Ultrafast light pulses can modify the electronic properties of quantum materials by perturbing the underlying, intertwined degrees of freedom. In particular, iron-based superconductors exhibit a strong coupling among electronic nematic fluctuations, spins, and the lattice, serving as a playground for ultrafast manipulation. Here we use time-resolved x-ray scattering to measure the lattice dynamics of photo-excited BaFe$_2$As$_2$. Upon optical excitation, no signature of an ultrafast change of the crystal symmetry is observed, but the lattice oscillates rapidly in time due to the coherent excitation of an $A_{1g}$~mode that modulates the Fe-As-Fe bond angle. We directly quantify the coherent lattice dynamics and show that even a small photo-induced lattice distortion can induce notable changes in the electronic and magnetic properties. Our analysis implies that transient structural modification can generally be an effective tool for manipulating the electronic properties of multi-orbital systems, where electronic instabilities are sensitive to the orbital character of bands near the Fermi level.}\\


\noindent One of the goals in materials research is to control quantum phases that emerge in strongly correlated materials, such as superconductivity and magnetism, since many of them exhibit exotic properties that promise applications in technology\cite{dagotto05}. While the microscopic mechanism of such emergence remains elusive, it is generally agreed that the formation and competition of quantum phases results from a subtle balance among the strongly coupled spin, charge, lattice and orbital degrees of freedom. Shifting this balance provides a promising avenue to manipulate emergent phenomena in strongly correlated materials.

In equilibrium the electronic properties are typically modified by chemical doping or application of an external parameter, e.g., magnetic fields, strain or hydrostatic pressure\cite{kimber09,nandi10,paglione10}. However, perturbing the subtle balance of interactions by using ultrafast light pulses to manipulate material properties in non-equilibrium transient states has recently received significant attention. Many studies, including the generation of coherent collective oscillatory states\cite{merlin97,sokolowski03,schmitt08,kubacka14} and transiently induced phases which have no analog in thermal equilibrium\cite{rini07,fausti11,kim12}, have demonstrated the power of these techniques. To date, most information about these photo-induced states is obtained by optical or photoemission spectroscopy, providing only limited and indirect insight on the dynamics of the lattice degree of freedom. Therefore, it is important to directly probe the complementary structural dynamics of these photo-induced states via time-resolved x-ray scattering with femtosecond resolution.

BaFe$_2$As$_2$, a parent compound of the high-temperature superconducting iron pnictides\cite{paglione10,johnston10}, is an ideal system for manipulating electronic properties via transient structural modification, as the lattice couples strongly to the magnetic and electronic degrees of freedom. Upon cooling, the system first undergoes a structural phase transition ($T_{\rm s}$), followed by a spin-density-wave (SDW) transition\cite{rotter08,kim11} at $T_{\rm N}$, just 0.75~K below $T_{\rm s}$. Importantly, the existence of nematic electronic fluctuations has been demonstrated at even higher temperatures\cite{chu10,chu12,boehmer14,fernandes14}; and their divergence drives the aforementioned structural phase transition. The electronic structure of pnictides also appears to be extremely sensitive to the Fe-As-Fe bond angle~$\alpha$ (Fig.~1a) as it changes the hybridization of the iron 3$d$ and arsenic 4$p$ orbitals---evidenced by band structure calculations\cite{kuroki09,lee14}---and correlates with the superconducting transition temperature in doped compounds\cite{kreyssig08,zhao08} as well as magnetism\cite{zhang14}. Notably, transient optical reflectivity\cite{mansart09}, conductivity\cite{kim12}, and time- and angle-resolved photoemission spectroscopy\cite{avigo13,yang14} (trARPES) revealed that an ultrafast optical excitation induces coherent oscillations with a frequency $f=5.45$~THz, corresponding to an $A_{1g}$ phonon mode observed in Raman spectroscopy\cite{rahlenbeck09}. Intriguingly, THz~spectroscopy\cite{kim12} indicates that exciting the coherent $A_{1g}$ phonon mode enhances magnetism by inducing a transient SDW state even above $T_{\rm N}$; and trARPES\cite{avigo13,yang14} finds concomitant strong modulations of the density of states near the Fermi level for similar excitations. However, disentangling the lattice's influence requires a direct structural characterization in the photo-induced transient state, which serves as an important experimental boundary condition for the associated variation of the electronic and magnetic degrees of freedom.

We employ time-resolved x-ray scattering at the Linac Coherent Light Source (LCLS), an x-ray free electron laser (FEL), to directly measure the photo-excited lattice dynamics in BaFe$_2$As$_2$. We map the temporal evolution of the crystal structure by recording the diffraction pattern at different time delays~$\Delta t$ between an 800 nm infrared (IR) pump pulse and the 8.7 keV x-ray probe pulse (Fig.~1b). Recently, Rettig \textit{ et al.}\cite{rettig14} conducted a similar study. Here we corroborate their findings, while also illuminating different experimental aspects and elucidating the impact of the lattice dynamics on the electronic and magnetic properties. In particular, we investigate two questions: (i)~Can the ultrafast photo-excitation trigger an ultrafast change of the crystal symmetry by perturbing the electronic nematic state? (ii)~How are the 5.45~THz coherent oscillations, as seen in both optical and photoemission spectroscopy, reflected in the lattice degree of freedom, and what are the consequences on the electronic and 
magnetic properties?\\

\noindent\textbf{\large Results} 

\noindent\textbf{Photo-induced lattice dynamics below $T_{\rm s}$.}~~~Figures~2a-d show the temperature dependence of the (118)$_{\rm T}$ lattice Bragg peak (in tetragonal notation) near the structural ($T_{\rm s}$) and the antiferromagnetic phase transition ($T_{\rm N}$) during slow cooling from a nominal temperature of $T=140$ to 137~K. For $T>T_{\rm s}$ the crystal structure is tetragonal, yielding a single peak on the detector (Fig.~2a). Upon cooling, the (118)$_{\rm T}$ peak splits (Fig.~2b-c), as a consequence of the tetragonal to orthorhombic structural phase transition (space group: $I4/mmm\rightarrow Fmmm$). The detailed evolution of the peak-splitting near $T_{\rm s}$ and $T_{\rm N}$ is depicted in Fig.~2d: it occurs continuously for temperatures $T_{\rm s}>T>T_{\rm N}$, followed by a sudden jump at the SDW ordering temperature $T_{\rm N}$. This behaviour is equivalent to the results obtained in thermal equilibrium using a synchrotron x-ray source\cite{kim11}.

Upon photo-excitation via femtosecond optical pulses, hot electrons are generated, populating states above the Fermi level, which then decay through allowed electron-electron and electron-phonon scattering channels. These incoherent scattering processes should, in principle, weaken the nematic fluctuations, which may allow the crystal structure to recover the original four-fold symmetry, i.e., the tetragonal phase.

To test the aforementioned conjecture, Fig.~3a-b show the temporal evolution of the split (118)$_{\rm T}$ peaks for a temperature $T_{\rm s}>T>T_{\rm N}$, along a line cut on the area detector (indicated in the inset). As a function of time the two orthorhombic peaks neither merge nor come closer, showing no signature of any ultrafast structural change from orthorhombic to tetragonal symmetry within $\Delta t=4.5$~ps after photo-excitation. Also, no evidence is found for a change of the lattice parameters in the picosecond regime, as the profile of the Bragg peaks does not shift or broaden (Fig.~2b). Therefore, we conclude that a structural transition, a process that involves the movement of all atoms to eliminate the orthorhombic structural domains\cite{tanatar09} and also depends on the strain potential from the bulk material, does not occur in BaFe$_2$As$_2$ on these ultrafast time scales at an absorbed fluence of 2.9~mJ/cm$^2$---approximately half of the sample damage threshold observed in the experiment.\\

\noindent\textbf{Direct quantification of the coherent lattice dynamics.}~~~Careful examination of the diffraction pattern as a function of time reveals ultrafast lattice dynamics. As shown in Fig.~3c, the intensity of both split Bragg peaks exhibits a time-dependent modulation following photo-excitation, suggesting that the entire probed sample volume is in a coherent oscillatory state with a period of approximately 185~fs.

This coherent state is characterized further at a slightly elevated temperature $T>T_{\rm s}$, where the improved signal-to-noise ratio facilitates a quantitative analysis, as the scattered intensity is concentrated in one single Bragg peak. Figure~4 shows the temporal evolution of the (118)$_{\rm T}$ line cut (Fig.~4a) and the integrated counts on the area detector (Fig.~4b), normalized to the intensity before time zero. Most striking is the rise of the (118)$_{\rm T}$ Bragg peak intensity with a maximum at $\Delta t\sim130$~fs after photo-excitation. Moreover, coherent oscillations are resolved with a periodicity of 185~fs, as already evidenced in Fig.~3c. The Fourier transform (FT) (inset of Fig.~4b) and the background-subtracted diffracted intensity (Fig.~4c) yield an oscillation with $f=5.45(4)$~THz that coincides with the frequency of the $A_{1g}$ phonon mode, as measured by Raman spectroscopy\cite{rahlenbeck09}. This finding provides strong support that the coherent oscillations indeed can be attributed to the Fe-As-Fe bond angle mode.

To better understand and quantify the lattice dynamics associated with the coherent excitation of the $A_{1g}$~phonon, we have performed a structure factor calculation. Since the associated eigenmode involves only the vertical displacement of the arsenic atoms, the structural change can be parametrized by the Fe-As-Fe bond angle $\alpha$ (Fig.~1a). In the presence of the $A_{1g}$ bond angle mode the structure factor can be written as
\begin{equation}
	F_{hkl}(\alpha)=\sum_{n} f_n\cdot \exp[2\pi i \cdot \mathbf{G}_{hkl} \cdot \mathbf{r}_n(\alpha)],
\end{equation}
where $n$ indexes individual atoms in the unit cell, $f_n$ is the dispersion-corrected atomic scattering factor\cite{henke93}, $\mathbf{r}_n(\alpha)$ is the atomic position and $\mathbf{G}_{hkl}$ is the scattering vector.
The $\alpha$-dependent diffracted intensity is obtained from the relation $I_{hkl}(\alpha) \propto |F_{hkl}(\alpha)|^2$.

The calculated relative intensity change $I_{118}(\alpha)/I_{118,\rm eq}$ is shown in Fig.~5a. The signal clearly increases from its equilibrium value with decreasing $\alpha$. A comparison with the raw data in Fig.~4b reveals that the initial ultrafast increase of the (118)$_{\rm T}$ Bragg peak intensity is associated with an ultrafast decrease of the bond angle $\Delta\alpha_{\rm max}=-0.62(4)^\circ$. Figure~5b depicts the temporal evolution of the bond angle change $\Delta\alpha(t)$, as deduced from the raw data shown in Fig.~4b without deconvolution of the finite time resolution, revealing an $A_{1g}$ oscillation amplitude $\Delta\alpha_{\rm osc}=0.27(8)^\circ$ (averaged amplitude of the first three oscillations, the error is determined by the standard deviation), in addition to the initial decrease of $\alpha$. The magnitude of $\Delta\alpha_{\rm osc}$ is in agreement with the results obtained by Rettig \textit{ et al.}\cite{rettig14}, after taking into account the pump fluence and the time resolution of the probe pulse. For clarity, we note that in Ref.~\onlinecite{rettig14} the $A_{1g}$ mode is parametrized in terms of the Fe-As tetrahedral angle and not the Fe-As-Fe bond angle. The experimentally established temporal dependence of the bond angle provides direct input for a theoretical evaluation of the associated transient variation of the electronic and magnetic degrees of freedom in this coherent oscillatory state.\\

\noindent\textbf{Consequence on the electronic and magnetic properties.}~~~To assess the qualitative influence of the transient modification of the crystal structure on magnetism\cite{kim12}, we have carried out self-consistent Hartree-Fock mean-field calculations. We employ a five-orbital, tight-binding fit to the density functional theory-derived band structure\cite{graser09} of LaFeAsO, which shows a qualitative similarity to BaFe$_{2}$As$_{2}$ at low doping and for energies near the Fermi level\cite{miyake10}. This simplifies the discussion by restricting the calculations to two-dimensions. Throughout the analysis we reference to the one iron Brillouin zone (BZ) notation, which provides additional clarity when discussing the evolution of the band structure and Fermi surfaces as a function of $\alpha$. The magnetic moment, and hence the N\'eel temperature $T_{\mathrm{N}}$, is determined at the mean-field level for a multi-orbital electron-electron interaction with parameters tuned to stabilize a $\mathbf{Q}=(\pi,0)$ SDW with six electrons per site\cite{plonka13,yi14}. We assume that the pnictogen height, which controls the bond angle $\alpha$, primarily affects the band structure parameters associated with the $d_{xy}$ orbital: the nearest and next-nearest neighbour hopping integrals\cite{vildosola08,kuroki09}. While the pnictogen height also affects other parameters, these changes are shown to have more influence at higher binding energies and less on the band structure close to the Fermi level\cite{kuroki09,yang14}. To mimic the bond angle in BaFe$_{2}$As$_{2}$, we subtract 2.4$^\circ$ from the equilibrium value of $\alpha$ in LaFeAsO and extrapolate the intra-orbital $d_{xy}$ hopping integrals linearly over a range of $\alpha$ following the dependence determined from Ref.~\onlinecite{kuroki09}. We note that the derived band structure (Fig.~6a) qualitatively agrees with the known SDW-folded band structure\cite{graser10,yi11}.

Figure~6a shows the influence of $\Delta\alpha =-1.2^\circ$, on the band structure close to the Fermi level in the SDW $(\pi,0)$-folded zone. Principally, the change in $\alpha$ raises the dominant $d_{xy}$~hole-band near the Y-point (folded from the one iron BZ M-point), consistent with the change in $d_{xy}$ hopping integrals\cite{kuroki09}. To maintain a consistent filling fraction, a rigid chemical potential shift leads to a lowering of the $d_{xz}$ and $d_{yz}$ bands at the $\Gamma$-point, which is consistent with trARPES\cite{avigo13,yang14}, and not connected to doping evolution of the equilibrium state. The significance of these changes becomes apparent when viewed on the Fermi surface (inset of Fig.~6a). Figures~6b-c depict the Fermi surface in the vicinity of the $\Gamma$- and Y-points in the SDW-folded BZ. A reduction of the Fe-As-Fe bond angle improves nesting at the Y-point considerably, where hole- and electron-bands of $d_{xy}$ character interact. While the change reduces nesting at the $\Gamma$-point, this has less impact on magnetism due the incompatibility of the orbital character on the bands, which already suppresses the opening of a SDW gap there. These effects are borne out by the change in the calculated N\'eel temperature as a function of $\alpha$, shown in Fig.~6d.\\

\noindent\textbf{\large Discussion} 

\noindent In general, our results highlight that coherent excitation of an optical phonon may allow manipulation of the electronic properties of multi-orbital systems, in which orbital physics is central to the electronic structure. In such compounds, electronic instabilities are driven by band edges with different orbital character close to the Fermi level, which are sensitive to small changes of the underlying crystal structure. Remarkably, Fig.~6d shows that a 0.6~\% change in $\alpha$, as induced by photo-excitation, results in a substantial enhancement of $\sim6.5$~\% in the calculated SDW transition temperature ($\Delta T_{\rm N}\sim9$~K) due to precisely these effects. This is qualitatively consistent with the recent observation\cite{kim12} of photo-induced transient SDW order at temperatures above $T_{\rm N}$. Although the onset of SDW order and the change in crystal symmetry are coupled in equilibrium, such a coupling may not hold in the photo-excited non-equilibrium state which may depend on the nature of the transient SDW order, e.g. fluctuating or static, and associated time scales. We note that the mean-field calculations do not include effects of fluctuations, which are crucial for short-range SDW correlations; and while the transient Fermi surface topology favours the emergence of SDW order, it does not take into account scattering processes due to the relaxation of photo-excited ``hot'' electrons. 

Nevertheless, one can already envision some exciting possibilities. For example, our model\cite{graser09} suggests a transient Lifschitz transition, the induction of an additional $d_{xy}$ pocket at the zone boundary, in LaFeAsO\cite{yang10}, LaFePO\cite{lu08} and other related systems, if $\alpha$ changes on the order of a degree via photo-excitation. Given the sizeable effect on the electronic and magnetic properties, it would be tantalizing to investigate how this transient coherent oscillatory state affects superconductivity in doped compounds. Theoretical and experimental studies already suggest an intimate connection between the Fe-As-Fe bond angle, the superconducting transition temperature\cite{kreyssig08,zhao08}, and the symmetry of the superconducting order parameter\cite{kuroki09}.\\

\small
\noindent\textbf{Methods}\\ The BaFe$_2$As$_2$ single-crystal was grown from self-flux and was of a millimetre size. It had a plate-like shape with the tetragonal $c$-axis perpendicular to the scattering surface that was prepared by cleaving. The lattice dynamics of the photo-excited single-crystal were studied at the X-ray Pump Probe (XPP) instrument of the Linac Coherent Light Source (LCLS) x-ray free electron laser\cite{emma10} at the SLAC National Accelerator Laboratory, benefiting from superb time resolution and x-ray pulse intensity. A dedicated sample chamber was assembled, allowing for low-temperature pump-probe hard x-ray scattering. All data reported here were measured at nominal temperatures $T=137-140$~K. The BaFe$_2$As$_2$ single-crystal was excited with an optical pump pulse and, thereafter, probed by a hard x-ray pulse. Both were operated with a repetition rate of 120~Hz.

The pump laser provided a $p$-polarized 800~nm IR pulse with a duration of $\sim55$~fs. The angle of incidence was $2^\circ$ with a spot size of 65 x 80 $\mu$m$^2$ ($h$ x $v$, Gaussian FWHM), yielding an absorbed fluence of $2.9$~mJ/cm$^2$. As the probe, $p$-polarized $E= 8.7$~keV x-rays from a silicon (111) monochromator, with a pulse duration of $\sim45$~fs, were used, resulting in a flux of $\sim10^{10}$~photons per pulse on the sample---well below the damage threshold. A combination of upstream slits and beryllium compound refractive lenses shaped the x-ray beam to 15~x~30~$\mu$m$^2$, in order to fit the photo-excited sample volume at 0.5$^\circ$ grazing incidence. 8.7 keV x-rays were used to match the penetration depths of the pump and probe pulses. The arrival time between the pump laser and the x-rays was measured pulse by pulse to allow for time-sorting\cite{lemke13} that mitigates the intrinsic jitter of the FEL and yields an overall time resolution of better than 75~fs. The x-ray diffraction patterns were recorded using a CSPAD-140k detector\cite{herrmann13} at full beam rate.

The Python package \textit{periodictable~1.4.1} was used to compute the dispersion-corrected atomic scattering factors\cite{henke93} in the structure factor calculation.\\

\vspace{10mm}

\noindent\textbf{Acknowledgements}\\This work was carried out at the X-ray Pump Probe (XPP) instrument of the Linac Coherent Light Source (LCLS) at the SLAC National Accelerator Laboratory. LCLS is an Office of Science User Facility operated for the U.S. Department of Energy, Office of Science by Stanford University. The authors gratefully acknowledge assistance and discussions with J.~J.~Turner, I.~R.~Fisher, J.-H.~Chu and H.-H.~Kuo. The research was supported by the U.S. Department of Energy, Office of Basic Energy Sciences, Division of Materials Sciences and Engineering under contract no. \mbox{DE-AC02-76SF00515}. S.G. and D.L. acknowledge support by the Swiss National Science Foundation under Fellowships No.~P2EZP2\_148737 and P300P2\_151328, respectively. K.W.K. was supported by the research grant of Chungbuk National University in 2012, Basic Science Research Program through the National Research Foundation of Korea (NRF) funded by the Ministry of  Science, ICT and Future Planning (2014R1A1A1007531) and PAL, Korea.\\

\noindent\textbf{Author contribution}\\W.S.L. and K.W.K. conceived the project with input from T.P.D. and Z.X.S. W.S.L., K.W.K., Y.Z., D.Z., M.Y., G.L.D., P.S.K., R.G.M., M.C., J.M.G., Y.F., J.S.L., A.M., Y.D.C., Z.H. and C.C.K. prepared the experiment and carried out the measurements. T.W. synthesized and characterized the single-crystal. S.G., Y.Z. and D.L. analysed the data. N.P., A.F.K., B.M. and T.P.D. carried out the theoretical evaluation. S.G. and W.S.L. wrote the manuscript with contributions from all co-authors.\\

\noindent\textbf{Competing financial interests}\\The authors declare no competing financial interests.\\

\clearpage

\noindent Figure 1:~~\textbf{Crystal structure and time-resolved x-ray scattering.} (\textbf{a}) Tetragonal crystal structure of BaFe$_2$As$_2$ in the presence of the $A_{1g}$ phonon mode, parametrized by the Fe-As-Fe bond angle $\alpha$. (\textbf{b}) Schematic of the experimental setup with the incoming optical pump (red) and the x-ray probe pulse (blue). The temporal evolution of the diffraction pattern from the photo-excited BaFe$_2$As$_2$ single-crystal was measured with a CSPAD-140k area detector. $\Delta t$ is the time delay of the probe pulse with respect to the pump pulse.\\

\noindent Figure 2:~~\textbf{Structural phase transitions without optical pumping.} (\textbf{a})-(\textbf{c}) Diffraction pattern of the (118)$_{\rm T}$ lattice Bragg peak at temperatures in the vicinity of the structural ($T_{\rm s}$) and magnetic ($T_{\rm N}$) phase transition. (\textbf{d}) Line cut on the area detector [dashed line in (\textbf{a})] slowly cooling from a nominal temperature of $T=140$ to 137~K. The tetragonal (118)$_{\rm T}$ Bragg peak splits first at $T_{\rm s}$, due to the transition to the orthorhombic crystal structure, and then further at $T_{\rm N}$ as a result of the onset of SDW order.\\

\noindent Figure 3:~~\textbf{Photo-induced lattice dynamics below $T_{\rm s}$.} (\textbf{a}) Temporal evolution of the line cut through the split Bragg peak at $T_{\rm s}>T>T_{\rm N}$ at an absorbed pump fluence of $2.9$~mJ/cm$^2$. The inset depicts the line cut on the area detector. (\textbf{b}) Diffraction peak profiles along the line cut at selected delay times. No changes are observed in peak position and width. (\textbf{c}) Subtraction of the averaged line cuts before time zero ($\Delta t = -0.5$ to 0 ps) reveals a photo-induced periodic intensity modulation of both orthorhombic domains for positive time delays.\\

\noindent Figure 4:~~\textbf{Photo-induced coherent lattice dynamics for $T>T_{\rm s}$.} (\textbf{a}) Temporal evolution of the line cut after subtraction of the averaged line cuts before time zero. (\textbf{b}) Integrated intensity on the area detector as a function of time. Both (\textbf{a}) and (\textbf{b}) show a distinct modulation of the (118)$_{\rm T}$ Bragg peak intensity after photo-excitation. Time zero ($\Delta t=0$) is defined as the time delay, at which one observes a rise of the diffracted Bragg peak intensity. The background (black line) is modelled as a convolution of the overall time resolution and an exponential decay of the initial ultrafast intensity rise, on a linear slope. The inset of (\textbf{b}) shows the Fourier transform of the background-subtracted integrated intensity (\textbf{c}), which both identify the coherent oscillations with the 5.45 THz $A_{1g}$ phonon mode.\\

\noindent Figure 5:~~\textbf{Dynamics of the Fe-As-Fe bond angle.} (\textbf{a}) Dependence of the (118)$_{\rm T}$ Bragg peak intensity on the bond angle $\alpha$ from a structure factor calculation. The shaded area indicates the magnitude of the initial change $\Delta\alpha_{\rm max}=-0.62(4)^\circ$, as obtained by comparison with the maximal intensity change of the integrated Bragg peak intensity in Fig.~4b. (\textbf{b}) The deduced temporal evolution $\Delta\alpha(t)$ from the raw data (without deconvolution of the finite time resolution), reveals an $A_{1g}$ oscillation amplitude $\Delta\alpha_{\rm osc}=0.27(8)^\circ$, following the initial decrease of $\alpha$.\\

\noindent Figure 6:~~\textbf{Influence of $\alpha$ on the electronic structure and SDW order.} (\textbf{a}) Effect of $\Delta\alpha=-1.2^\circ$ on the low-energy bands along the $\Gamma$-Y momentum cut in the SDW-folded BZ. The equilibrium bands (solid lines) shift as a result of the change in the Fe-As-Fe bond angle (dashed lines). The dominant ($>50\%$) $d$-orbital character for each band is colour-coded. We choose twice the experimentally observed $\Delta\alpha_{\rm max}$ to better illustrate the qualitative change. The inset shows the equilibrium Fermi surface and the locations of the $\Gamma$ and Y points. The unshaded area represents the $\mathbf{Q}=(\pi,0)$ SDW-folded BZ. The squares enclose portions that are enlarged in panels (\textbf{b}) and (\textbf{c}) to show the effect of $\Delta\alpha=-1.2^\circ$ on the Fermi surface in the SDW-folded BZ. Equilibrium Fermi surface pockets (left half of each panel) shift to new positions under the change of $\alpha$ (right half of each panel). Improved nesting of bands with similar orbital character ($d_{xy}$) is observed at the Y-point. (\textbf{c}) Results of self-consistent Hartree-Fock mean-field calculations for the relative change of $T_{\rm N}$ as a function of $\Delta\alpha$. The arrow indicates a $6.5\%$ increase in $T_{\rm N}$ for the experimentally observed $\Delta\alpha_\mathrm{max}\approx-0.6^\circ$. $\alpha_\mathrm{tet}$ is the Fe-As-Fe bond angle for a regular FeAs$_4$ tetrahedron, where superconductivity is found to be maximal in iron-based compounds\cite{kreyssig08,zhao08}. The line is a fit through the full data set as obtained from the mean-field calculations.\\

\begin{figure}[h]
	\includegraphics[width=0.55\linewidth]{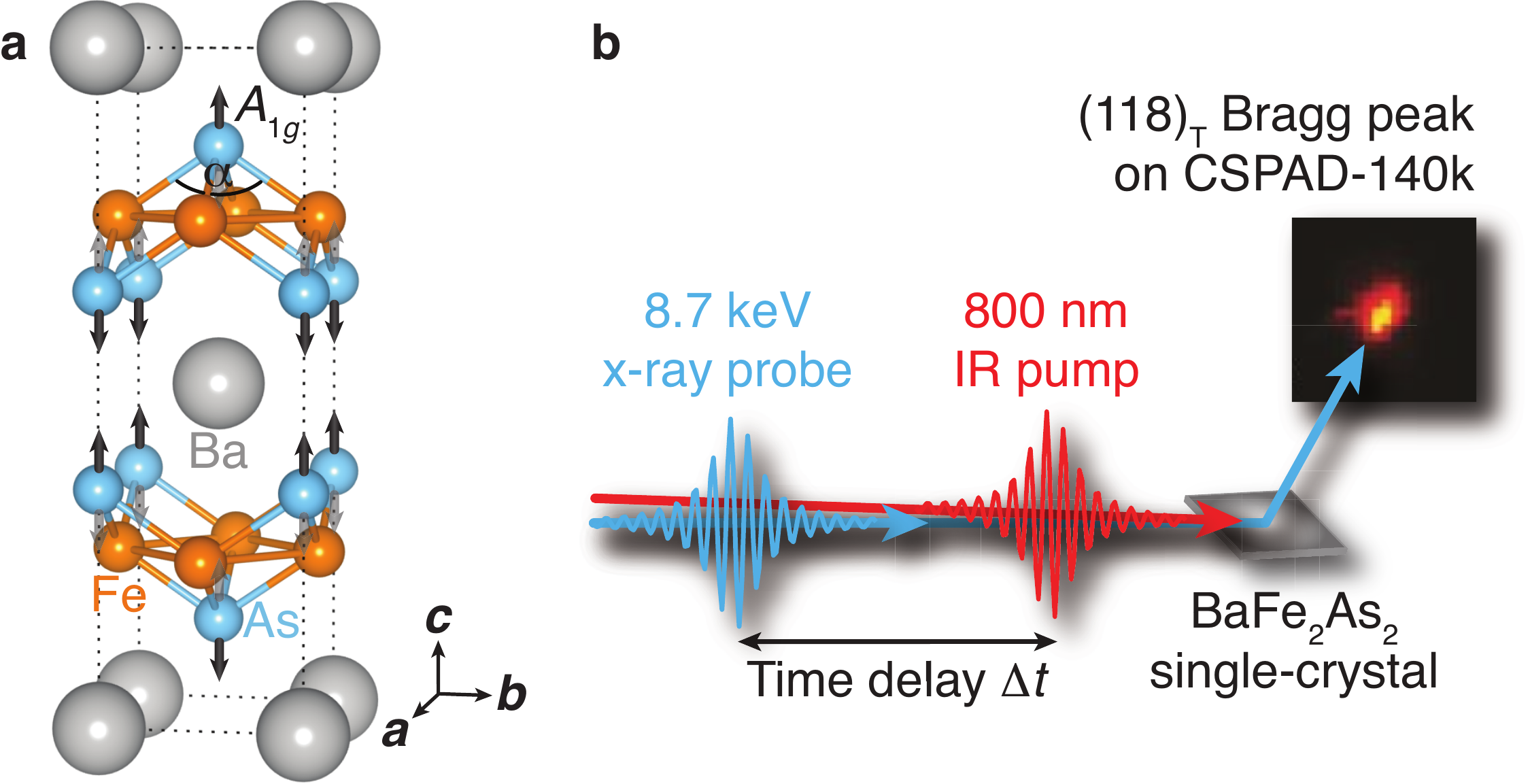}
	\caption{Crystal structure and time-resolved x-ray scattering.}
	\vspace{15mm}
\end{figure}

\begin{figure}
	\includegraphics[width=0.48\linewidth]{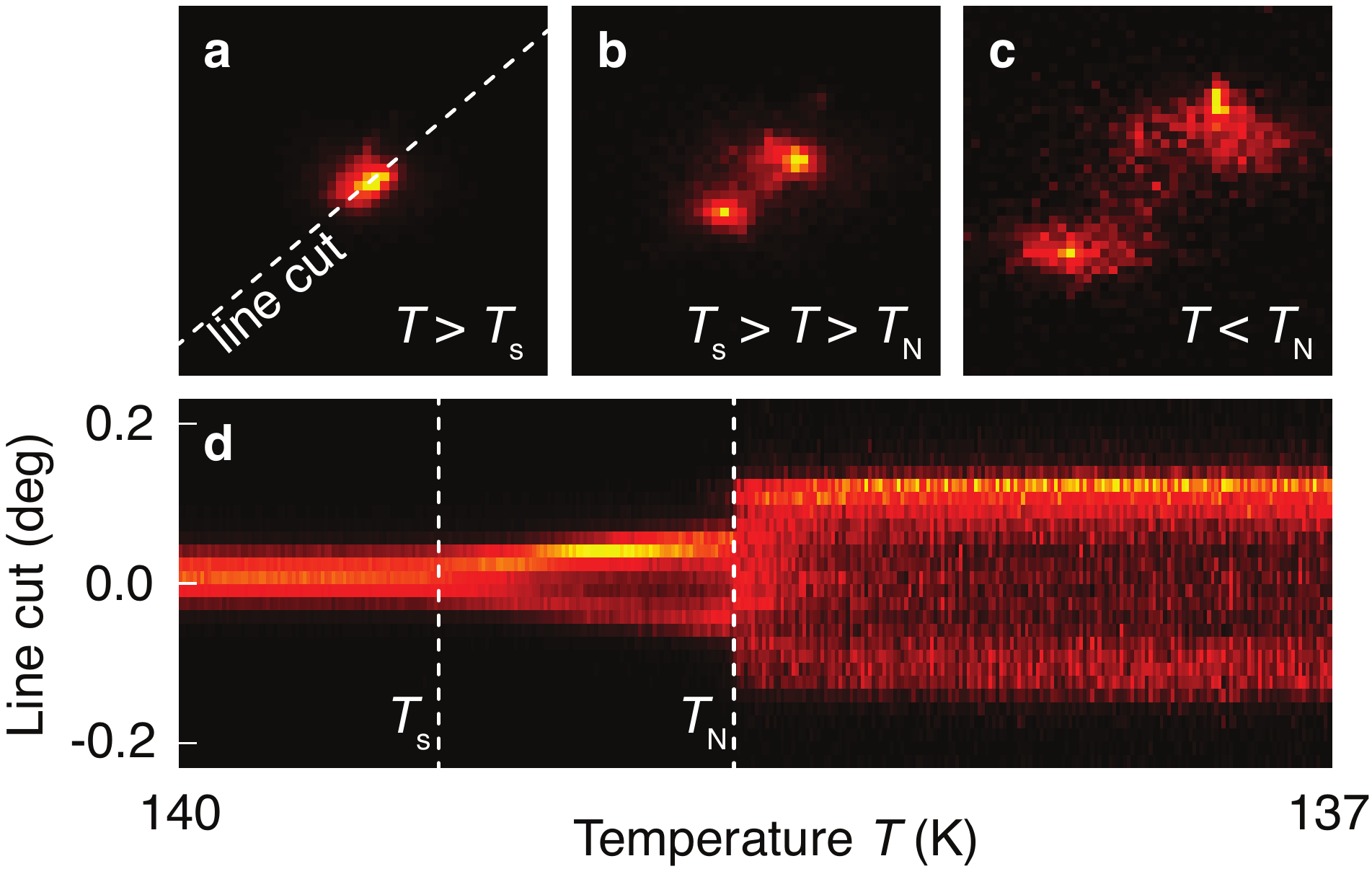}
	\caption{Structural phase transitions without optical pumping.}
	\vspace{15mm}
\end{figure}

\begin{figure}
	\includegraphics[width=0.53\linewidth]{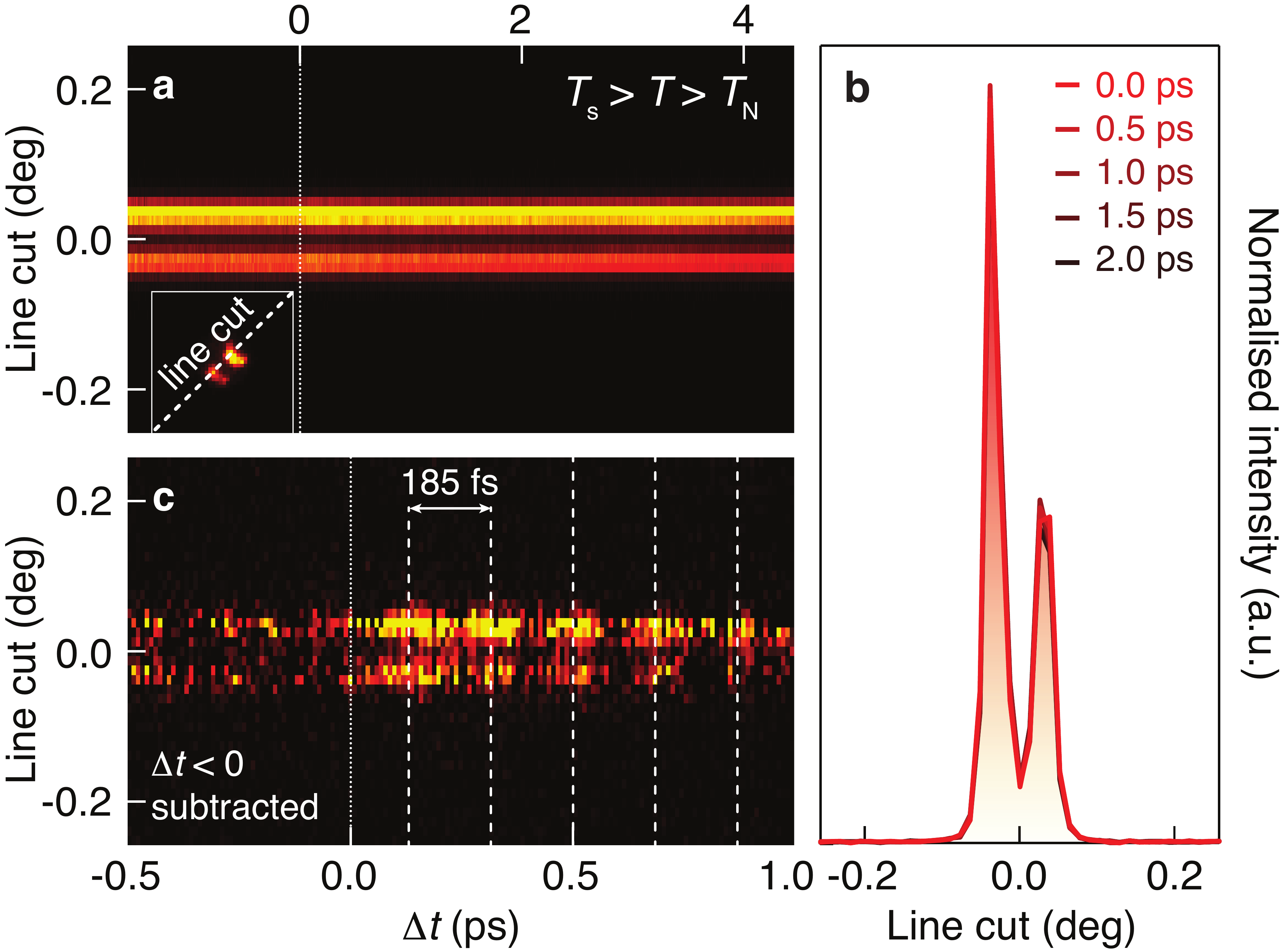}
	\caption{Photo-induced lattice dynamics below $T_{\rm s}$.}
	\vspace{15mm}
\end{figure}

\begin{figure}
	\includegraphics[width=0.51\linewidth]{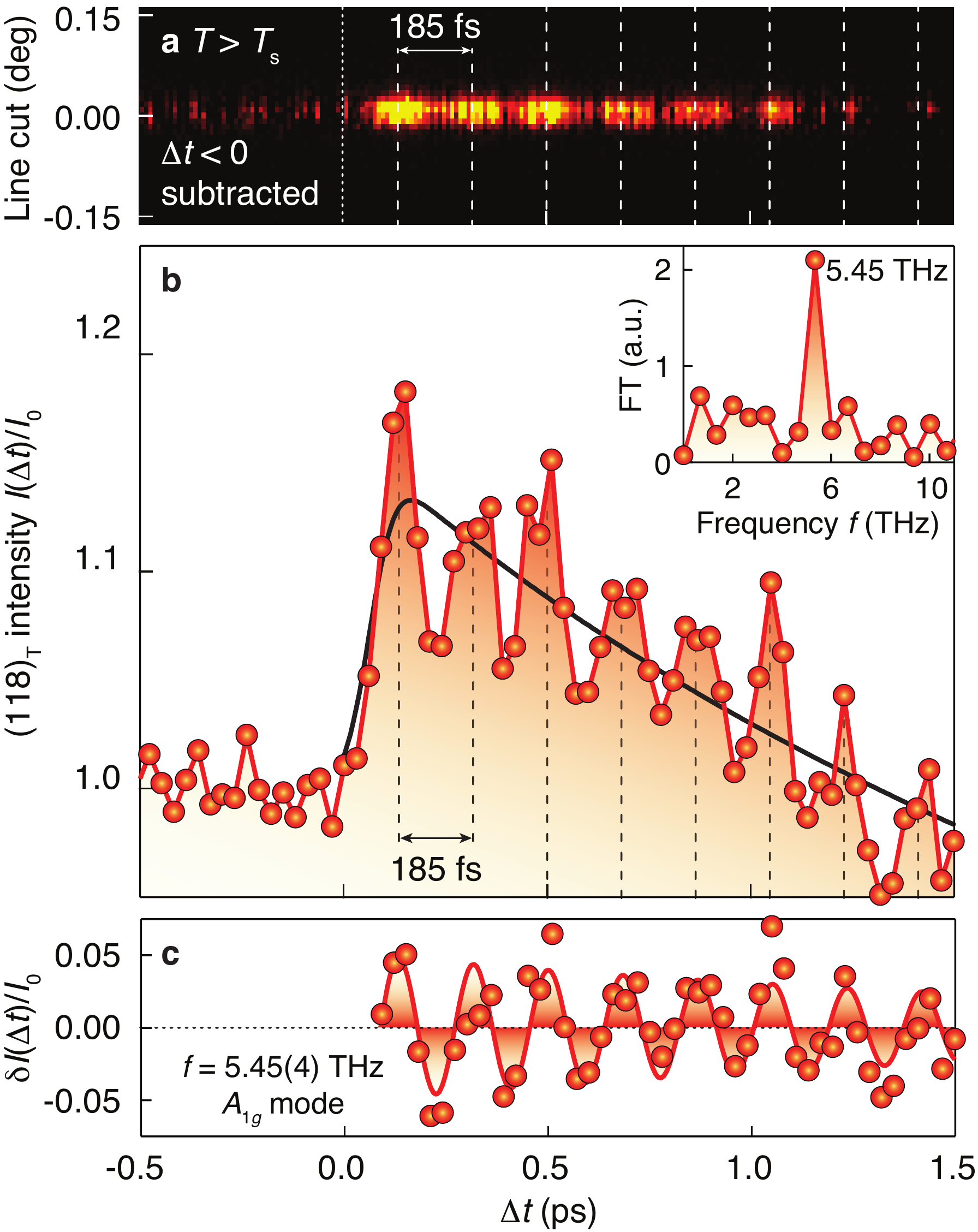}
	\caption{Photo-induced coherent lattice dynamics for $T>T_{\rm s}$.}
	\vspace{15mm}
\end{figure}

\begin{figure}
	\includegraphics[width=0.52\linewidth]{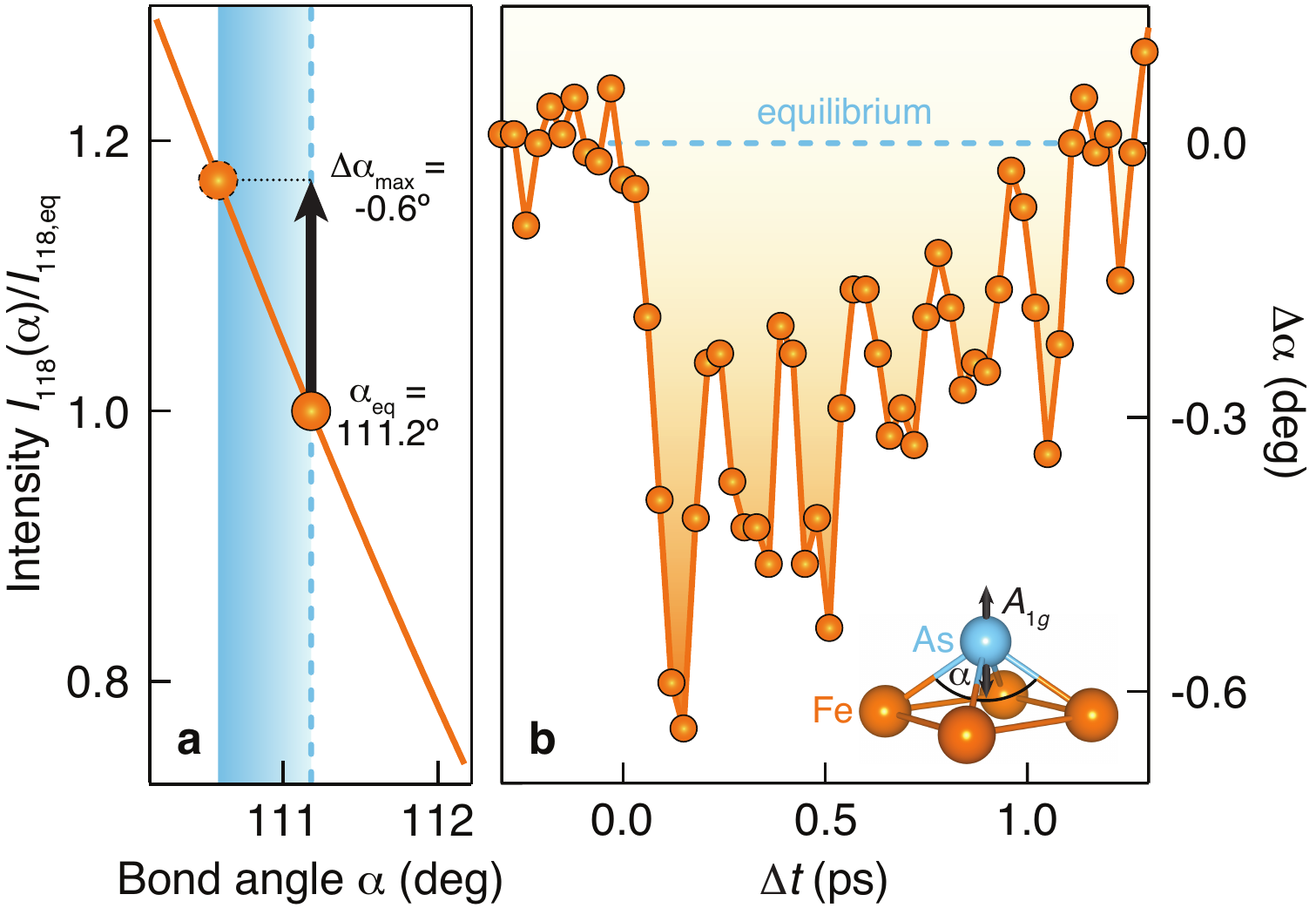}
	\caption{Dynamics of the Fe-As-Fe bond angle.}
	\vspace{15mm}
\end{figure}

\begin{figure}
	\includegraphics[width=\linewidth]{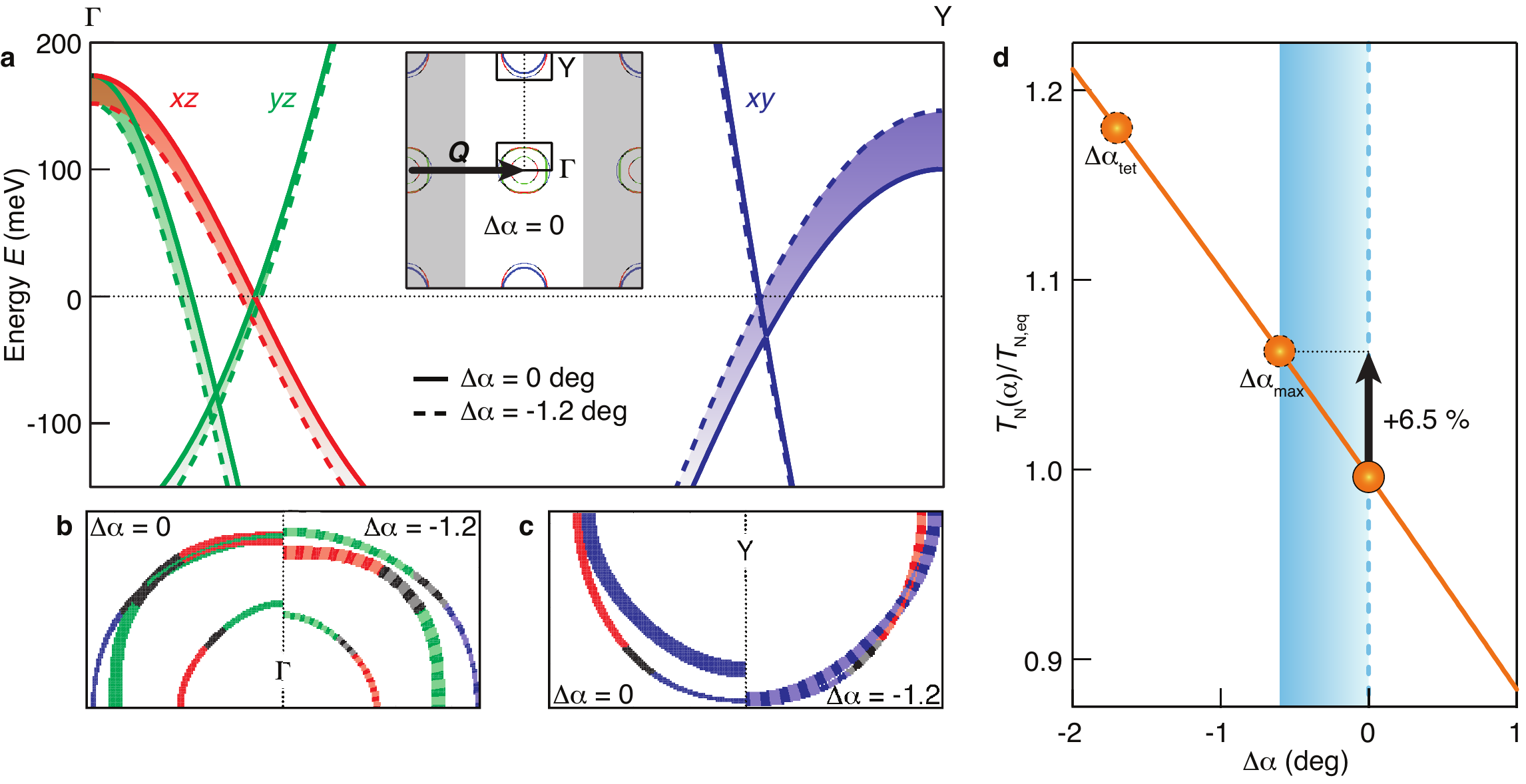}
	\caption{Influence of $\alpha$ on the electronic structure and SDW order.}
\end{figure}


\begin{thebibliography}{100} 

\bibitem{dagotto05} Dagotto, E. Complexity in strongly correlated electronic systems. \textit{Science} \textbf{309,} 257--262 (2005).

\bibitem{kimber09} Kimber, S. A. J. \textit{et~al.} Similarities between structural distortions under pressure and chemical doping in superconducting BaFe$_2$As$_2$. \textit{Nature Mater.} \textbf{8,} 471--475 (2009).

\bibitem{nandi10} Nandi, S. \textit{et~al.} Anomalous suppression of the orthorhombic lattice distortion in superconducting Ba(Fe$_{1-x}$Co$_x$)$_2$As$_2$ single crystals. \textit{Phys. Rev. Lett.} \textbf{104,} 057006 (2010).

\bibitem{paglione10} Paglione, J. \& Greene, R. L. High-temperature superconductivity in iron-based materials. \textit{Nature Phys.} \textbf{6,} 645--658 (2010).

\bibitem{merlin97} Merlin, R. Generating coherent THz phonons with light pulses. \textit{Solid State Commun.} \textbf{102,} 207--220 (1997).

\bibitem{sokolowski03} Sokolowski-Tinten, K. \textit{et al.}. Femtosecond x-ray measurement of coherent lattice vibrations near the Lindemann stability limit. \textit{Nature} \textbf{422,} 287--289 (2003).

\bibitem{schmitt08} Schmitt, F. \textit{et al.} Transient electronic structure and melting of a charge density wave in TbTe$_3$. \textit{Science} \textbf{321,} 1649--1552 (2008).

\bibitem{kubacka14} Kubacka, T. \textit{et al.} Large-amplitude spin dynamics
driven by a THz pulse in resonance with an electromagnon. \textit{Science} \textbf{343,} 1333--1336 (2014).

\bibitem{rini07} Rini, M. \textit{et al.} Control of the electronic phase of a manganite by mode-selective vibrational excitation. \textit{Nature} \textbf{449,} 72--74 (2007).

\bibitem{fausti11} Fausti, D. \textit{et al.} Light-induced superconductivity in a stripe-ordered cuprate. \textit{Science} \textbf{331,} 189--191 (2011).

\bibitem{kim12} Kim, K. W. \textit{et al.} Ultrafast transient generation of spin-density-wave order in the normal state of BaFe$_2$As$_2$ driven by coherent lattice vibrations. \textit{Nature Mater.} \textbf{11,} 497--501 (2012).

\bibitem{johnston10} Johnston, D. C. The puzzle of high temperature superconductivity in layered iron pnictides and chalcogenides. \textit{Adv. Phys.} \textbf{59,} 803--1061 (2010).

\bibitem{rotter08} Rotter, M. \textit{et al.} Spin-density-wave anomaly at 140 K in the ternary iron arsenide BaFe$_2$As$_2$. \textit{Phys. Rev. B} \textbf{78,} 020503(R) (2008).

\bibitem{kim11} Kim, M. G. \textit{et al.} Character of the structural and magnetic phase transitions in the parent and electron-doped BaFe$_2$As$_2$ compounds. \textit{Phys. Rev. B} \textbf{83,} 134522 (2011).

\bibitem{chu10} Chu, J.-H. \textit{et al.} In-plane resistivity anisotropy in an underdoped iron arsenide superconductor. \textit{Science} \textbf{329,} 824--826 (2010).

\bibitem{chu12} Chu, J.-H., Kuo, H.-H., Analytis, J. G. \& Fisher, I. R. Divergent nematic susceptibility in an iron arsenide superconductor. \textit{Science} \textbf{337,} 710--712 (2012).

\bibitem{boehmer14} B\"oehmer, A. E. \textit{et al.} Nematic susceptibility of hole-doped and electron-doped BaFe$_2$As$_2$ iron-based superconductors from shear modulus measurements. \textit{Phys. Rev. Lett.} \textbf{112,} 047001 (2014).

\bibitem{fernandes14} Fernandes, R. M., Chubukov, A.~V. \& Schmalian, J. What drives nematic order in iron-based superconductors? \textit{Nature Phys.} \textbf{10,} 97--104 (2014).

\bibitem{kuroki09} Kuroki, K., Usui, H., Onari, S., Arita, R. \& Aoki, H. Pnictogen height as a possible switch between high-$T_{\rm c}$ nodeless and low-$T_{\rm c}$ nodal pairings in the iron-based superconductors. \textit{Phys. Rev. B} \textbf{79,} 224511 (2009).

\bibitem{lee14} Lee, J. D., Yun, W. S. \& Hong, S. C. Ultrafast above-transition-temperature resurrection of spin density wave driven by coherent phonon generation in BaFe$_2$As$_2$. \textit{New J. Phys.} \textbf{16,} 043010 (2014).

\bibitem{kreyssig08} Kreyssig, A. \textit{et al.} Pressure-induced volume-collapsed tetragonal phase of CaFe$_2$As$_2$ as seen via neutron scattering. \textit{Phys. Rev. B} \textbf{78,} 184517 (2008).

\bibitem{zhao08} Zhao, J. \textit{et al.} Structural and magnetic phase diagram of CeFeAsO$_{1-x}$F$_x$ and its relation to high-temperature superconductivity. \textit{Nature Mater.} \textbf{7,} 953--959 (2008).

\bibitem{zhang14} Zhang, C. \textit{et al.} Effect of pnictogen height on spin waves in iron pnictides. \textit{Phys. Rev. Lett.} \textbf{112,} 217202 (2014).

\bibitem{mansart09} Mansart, B. \textit{et al.} Observation of a coherent optical phonon in the iron pnictide superconductor Ba(Fe$_{1-x}$Co$_x$)$_2$As$_2$ ($x=0.06$ and 0.08). \textit{Phys. Rev. B} \textbf{80,} 172504 (2009).

\bibitem{avigo13} Avigo, I. \textit{et al.} Coherent excitations and electron-phonon coupling in Ba/EuFe$_2$As$_2$ compounds investigated by femtosecond time- and angle-resolved photoemission spectroscopy. \textit{J. Phys.: Condens. Matter} \textbf{25,} 094003 (2013).

\bibitem{yang14} Yang, L. X. \textit{et al.} Ultrafast modulation of the chemical potential in BaFe$_2$As$_2$ by coherent phonons. \textit{Phys. Rev. Lett.} \textbf{112,} 207001 (2014).

\bibitem{rahlenbeck09} Rahlenbeck, M. \textit{et al.} Phonon anomalies in pure and underdoped $R_{1-x}$K$_x$Fe$_2$As$_2$ ($R=$~Ba, Sr) investigated by Raman
light scattering. \textit{Phys. Rev. B} \textbf{80,} 064509 (2009).

\bibitem{rettig14} Rettig, L. \textit{et al.} Ultrafast structural dynamics of the Fe-pnictide parent compound BaFe$_2$As$_2$. arXiv:1411.0718.

\bibitem{tanatar09} Tanatar, M. A. \textit{et al.} Direct imaging of the structural domains in the iron pnictides $A$Fe$_2$As$_2$ ($A=$~Ca, Sr, Ba). Phys. Rev. B \textbf{79,} 180508(R) (2009).

\bibitem{henke93} Henke, B. L., Gullikson, E. M. \& Davis, J. C. X-ray interactions: photoabsorption, scattering, transmission, and reflection at $E = 50 - 30,000$~eV, $Z = 1 - 92$. \textit{At. Data Nucl. Data Tables} \textbf{54,} 181--342 (1993).

\bibitem{graser09} Graser, S., Maier, T. A., Hirschfeld, P. J. \& Scalapino, D. J. Near-degeneracy of several pairing channels in multiorbital models for the Fe pnictides. \textit{New J. Phys.} \textbf{11,} 025016 (2009).

\bibitem{miyake10} Miyake, T., Nakamura, K., Arita, R. \& Imada, M. Comparison of ab initio low-energy models for LaFePO, LaFeAsO, BaFe$_2$As$_2$, LiFeAs, FeSe, and FeTe: electron correlation and covalency. \textit{J. Phys. Soc. Jpn.} \textbf{79,} 044705 (2010).

\bibitem{plonka13} Plonka, N., Kemper, A. F., Graser, S., Kampf, A. P. \& Devereaux, T. P. Tunneling spectroscopy for probing orbital anisotropy in iron pncitides. \textit{Phys. Rev. B} \textbf{88,} 174518 (2013).

\bibitem{yi14} Yi, M. \textit{et al.} Dynamic competition between spin-density wave
order and superconductivity in underdoped Ba$_{1-x}$K$_x$Fe$_2$As$_2$. \textit{Nat. Commun.} \textbf{5,} 3711 (2014).

\bibitem{vildosola08} Vildosola,V., Pourovskii, L., Arita, R., Biermann, S. \& Georges, A. Bandwidth and Fermi surface of iron oxypnictides: covalency and sensitivity
to structural changes. \textit{Phys. Rev. B} \textbf{78,} 064518 (2008).

\bibitem{graser10} Graser, S. \textit{et al.}. Spin fluctuations and superconductivity in a three-dimensional tight-binding model for BaFe$_2$As$_2$. \textit{Phys. Rev. B} \textbf{81,} 214503 (2010).

\bibitem{yi11} Yi, M. \textit{et al.} Symmetry-breaking orbital anisotropy observed for detwinned Ba(Fe$_{1-x}$Co$_x$)$_2$As$_2$ above the spin density wave transition. \textit{Proc. Natl. Acad. Sci.} \textbf{108,} 6878--6883 (2011).

\bibitem{yang10} Yang, L. X. \textit{et al.} Surface and bulk electronic structures of LaFeAsO studied by angle-resolved photoemission spectroscopy. \textit{Phys. Rev. B} \textbf{82,} 104519 (2010).

\bibitem{lu08} Lu, D. H. \textit{et al.} Electronic structure of the iron-based superconductor LaOFeP. \textit{Nature} \textbf{455,} 81--84 (2008).

\bibitem{emma10} Emma, P. \textit{et al.} First lasing and operation of an
\aa{}ngstrom-wavelength free-electron laser. \textit{Nature Photon.} \textbf{4,} 641--647 (2010).

\bibitem{lemke13} Lemke, H. T. \textit{et al.} Femtosecond optical/hard x-ray timing diagnostics at an FEL: implementation and performance. \textit{Proc. of SPIE} \textbf{8778,} 87780S (2013).

\bibitem{herrmann13} Herrmann, S. \textit{et al.} CSPAD-140k: A versatile detector for LCLS experiments. \textit{Nucl. Instrum. Methods A} \textbf{718,} 550--553 (2013).

\end{thebibliography}
\end{document}